%% file: main.tex
\documentclass[conference]{IEEEtran}

\usepackage{cite}

\usepackage{amsmath,amssymb,amsfonts}

\usepackage{algorithmic}

\usepackage{graphicx}
\usepackage{textcomp}
\usepackage{xcolor}

\usepackage[hyphens]{url}
\usepackage{fancyhdr}
\usepackage{hyperref}

\usepackage[LGRgreek, italic]{mathastext}
\usepackage{multirow,tabularx}
\usepackage{makecell}

\graphicspath{ {images/} }

\begin{document}
\title{An Energy-Efficient Near-Data Processing Accelerator for DNNs that Optimizes Data Accesses}
\author{\IEEEauthorblockN{Bahareh Khabbazan, Marc Riera, Antonio González }
\IEEEauthorblockA{\textit{dept. of Computer Architecture } \\
\textit{Universitat Polit\`{e}cnica de Catalunya (UPC)}\\
Barcelona, Spain\\
\{bahareh.khabbazan, marc.riera.villanueva, antonio.gonzalez\}@upc.edu}
}



\maketitle

\begin{abstract}
The constant growth of DNNs makes them challenging to implement and run efficiently on traditional compute-centric architectures. Some accelerators have attempted to add more compute units and on-chip buffers to solve the memory wall problem without much success, and sometimes even worsening the issue since more compute units also require higher memory bandwidth. Prior works have proposed the design of memory-centric architectures based on the Near-Data Processing (NDP) paradigm. NDP seeks to break the memory wall by moving the computations closer to the memory hierarchy, reducing the data movements and their cost as much as possible. The 3D-stacked memory is especially appealing for DNN accelerators due to its high-density/low-energy storage and near-memory computation capabilities to perform the DNN operations massively in parallel. However, memory accesses remain as the main bottleneck for running modern DNNs efficiently.

To improve the efficiency of DNN inference we present QeiHaN, a hardware accelerator that implements a 3D-stacked memory-centric weight storage scheme to take advantage of a logarithmic quantization of activations. In particular, since activations of FC and CONV layers of modern DNNs are commonly represented as powers of two with negative exponents, QeiHaN performs an implicit in-memory bit-shifting of the DNN weights to reduce memory activity. Only the meaningful bits of the weights required for the bit-shift operation are accessed. Overall, QeiHaN reduces memory accesses by 25\% compared to a standard memory organization. We evaluate QeiHaN on a popular set of DNNs. On average, QeiHaN provides $4.3x$ speedup and $3.5x$ energy savings over a Neurocube-like accelerator.
\end{abstract}

\begin{IEEEkeywords}
DNN, NDP, Accelerators, 3D-Stacked Memory, Quantization, Exponential, Transformer
\end{IEEEkeywords}

\input{sections/1-introduction}
\input{sections/2-related_research}
\input{sections/3-analysis}

\input{sections/4-proposed_method}
\input{sections/5-methodology}
\input{sections/6-evaluation}
\input{sections/7-conclusion}
\input{sections/8-acknowledgement}


\bibliographystyle{IEEEtranS} 
\bibliography{references}

\end{document}

%% file: sections/1-introduction.tex










\section{Introduction}
Deep Neural Networks (DNNs) represent the state-of-the-art solution to a broad range of machine learning applications such as natural language processing (NLP) and image classification. Modern DNNs can outperform human-level accuracy in many of these applications at the expense of high computational cost, memory requirements, and energy consumption. Complex DNN models are composed of hundreds of layers of artificial neurons with billions of model parameters and operations. The constant growth of DNNs makes them challenging to implement and run efficiently, even in the most recent accelerators~\cite{survey_accelerators} based on traditional computing architectures due to the memory wall problem. On the other hand, some recent research has focused on a new paradigm named Near-Data Processing (NDP)~\cite{oursurvey, processing}, which seeks to break the memory wall by moving the computations closer to the memory hierarchy.

Conventional DNN accelerators tend to dedicate a significant part of their area to the processing elements (PEs) that are responsible to speed-up the frequent dot-product operations of DNN layers. Due to the intrinsic parallel nature of DNN computations, many previous works aimed to exploit data and thread-level parallelism by having large PE arrays, what further stresses the memory bandwidth demands, which may become a main bottleneck that is unable to provide enough data to all PEs. The memory wall remains a major problem despite several attempts to tackle it by minimizing the off-chip memory accesses, maximizing the on-chip memory reuse factor~\cite{systematic_approach}, and increasing the on-chip buffer sizes in each PE. As a result, DNNs are heavily constrained by compute-centric architectures due to the high memory storage and memory bandwidth demands.

In addition, data movements normally represent the major cause of energy consumption. For instance, a recent study on Google workloads~\cite{google_workloads} shows that the data movements between memory and compute units contribute to 62.7\% of the total energy consumption. Consequently, the energy cost of the data transfers is orders of magnitude higher than that of the computations. Another observation in this study indicates that most of the data movements in consumer workloads are generated by simple functions and primitives that can be implemented in hardware at low cost. These observations, together with the effects of the memory wall and the dramatic increase in the size of DNNs, motivate the transition from conventional compute-centric to data-centric architectures for data-intensive applications.

Over the last few years, researchers have been exploring novel memory-centric architectures based on the so-called Near-Data Processing (NDP) paradigm to accelerate neural network algorithms by moving most of the computations ”in/near-memory” and, hence, reducing the data movements and their cost as much as possible. NDP has gained a lot of attention with the introduction of the 3D stack memory technology, which allows the integration of logic and memory in the same chip by stacking multiple dies vertically, providing high-speed connections between a high-density memory and a logic die. Micron’s Hybrid Memory Cube (HMC)~\cite{HMC}, High Bandwidth Memory (HBM)~\cite{HBM} from AMD/Hynix, and Samsung’s Wide I/O~\cite{Samsung} are popular examples implementing this trending technology.

NDP architectures based on 3D-stacked memory attack the memory wall by increasing storage capacity, memory bandwidth, and reducing power consumption~\cite{unison}. Compared with the conventional 2D DRAM, 3D memory provides an order of magnitude higher bandwidth (160 to 250 GBps) with up to $5x$ better energy efficiency and, hence, 3D memory is an excellent option for meeting the high throughput, low energy requirements of scalable DNN accelerators~\cite{new_DRAM_architecture, transpim, aga}. All 3D-stacked memory system implementations provide highly parallel access to memory which is well suited to the highly parallel architecture of the DNN accelerators~\cite{oursurvey, 2023abndp}.



Neurocube~\cite{neurocube} and TETRIS~\cite{tetris} are popular NDP 3D-stacked memory architectures that offer promising performance and energy consumption for accelerating DNNs. However, there is still large room for improvement, since these architectures and memory technology present multiple challenges to extend their adoption. First, architectures based on 3D-memory require to rethink of the design of on-chip buffers in the logic die as well as the location where the computations are executed. For example, performing simple operations on the DRAM dies can drastically reduce the amount of memory movements and the need for big on-chip buffers. Second, new approaches for dataflow scheduling and partitioning of the DNN computations are also required to reduce the memory pressure. Thus, changing the memory organization and data placement can fully exploit the features of 3D-stacked architectures. In addition, the area of the logic die is constrained by the package, and there are tight thermal constraints that limit the power dissipation of the system. Consequently, it is critical to propose solutions that improve in these aspects.


In this paper, we show how to efficiently exploit a logarithmic base-2 quantization (LOG2) of activations on FC and CONV layers of typical DNN models. First, we perform an analysis of the exponents obtained after the LOG2 quantization, and observed that a huge percentage of the activations are represented with negative exponents, that is, their original value is in the range of [-1, 1]. LOG2 quantization has been proposed in previous works to reduce the numerical precision of either activations/weights and exploited to substitute multiplications by a bit-shifting of the other operand. Based on these observations we propose an implicit in-memory bit-shifting of the DNN weights to reduce the memory movements. Weights are uniformly quantized and stored at the bit-level granularity into different memory regions, that is, each bit of a set of weights is stored in a different memory bank to exploit the inherent parallelism of 3D-stacked architectures. Next, we propose a mechanism to avoid accessing the bits of the weights that are not useful due to the right bit-shifting of the negative exponents of the logarithmically quantized activations.

Then, we present QeiHaN, a novel NDP accelerator that implements the above LOG2 quantization-shifting engine and efficient weight storage scheme for high-performance low-energy DNN inference. QeiHaN is implemented on top of a Neurocube-like architecture, but extended with an enhanced input stationary dataflow. The extra hardware required for our technique is modest since most of the components are already available in the baseline. QeiHaN only requires a small set of additional comparators and integer adders to perform the LOG2 quantization. Then, we also replace the multipliers by simple bit-shift logic, reducing the computational cost and the overall area of the PEs. Our experimental results show that the overheads are minimal compared to the savings in memory accesses and multiplications.

To summarize, this paper focuses on efficient DNN inference leveraging logarithmic quantization in NDP 3D-stacked DRAM-based accelerators. The main contributions are:

\begin{itemize}

\item We analyze the distribution of exponents of the logarithmically (i.e. LOG2) quantized activations in multiple layers of modern DNNs including CNNs, RNNs, and Transformers. We observe that a huge percentage of the exponents are negative, leading to potential memory savings as a result of reducing the accesses to only the useful bits of the weights.

\item We propose a novel data layout and an optimized data flow to exploit the bank-level parallelism of 3D-stacked memory together with the LOG2 quantization of activations. Each memory bank stores a different subset of the bits of the uniformly quantized weights to allow for parallel accesses to the required bits of the bit-shifting operations. On average, we reduce the memory accesses due to the weights by 25\% compared to a standard memory organization.

\item We present QeiHaN, a 3D-stacked DRAM-based hardware accelerator that implements our data layout and dataflow for efficient DNN inference. We evaluate QeiHaN for several DNNs. QeiHaN improves performance by $1.4x$ and reduces energy consumption by $1.3x$ on average over NaHiD, a baseline accelerator implementing the same dataflow and quantization as QeiHaN but with a standard memory organization for weights. Compared to Neurocube~\cite{neurocube}, QeiHaN achieves $4.3x$ speedup and $3.5x$ energy savings on average.

\end{itemize}

The rest of the paper is organized as follows. Section~\ref{Preliminaries} introduces some preliminaries for QeiHaN and provides a summary of works related to 3D memory DNN accelerators. Section~\ref{analysis} discusses the observations on the logarithmic quantization of activations for a modern set of DNNs. Section~\ref{QeiHaN Architecture} describes the architecture of QeiHaN including the implementation details of the main hardware components. Section~\ref{methodology} presents the evaluation methodology and Section~\ref{evaluation} discusses the experimental results of QeiHaN on different networks. Finally, Section~\ref{conclusion} concludes the paper by summarizing the key insights of this design alongside the overall performance.

%% file: sections/2-related_research.tex





\section{Background \& Related Work}\label{Preliminaries}
In the following subsections we review some terminology and concepts that may be helpful throughout this paper. First, we give a general description of DNNs, including the main categories and different types of layers. Next, we review DNN quantization and common dataflows of DNN accelerators. Finally, we discuss 3D memory architectures, which offer more opportunities to implement a highly efficient DNN accelerator in terms of both performance and energy consumption.

\subsection{Modern DNNs}\label{DNN_classification}
Deep Neural Networks (DNNs) are classified into three main categories. First, Multi-Layer Perceptrons (MLP) consist of multiple Fully-Connected (FC) layers in which every input neuron is connected, via synapses with particular weights, to every output neuron. Second, Convolutional Neural Networks (CNN) are composed of multiple convolutional layers to extract features, usually followed by one or several FC layers to perform the final classification. CNNs, such as AlexNet~\cite{alexnet}, have demonstrated to be particularly efficient for image and video processing. Finally, Recurrent Neural Networks (RNN)~\cite{recurrent} are made of multiple layers of cells with feedback connections, stacked on top of each other. RNN cells store information from past executions to improve the accuracy of future predictions. The most popular RNN architecture is the Long–Short Term Memory (LSTM) cell, which consists of multiple single-layer FC networks commonly referred as gates. PTBLM~\cite{ptblm}, an example of LSTM-based RNN, is used for various applications like language modeling, speech recognition, and machine translation.

Attention-based DNNs, such as the Transformer~\cite{attention} and all the BERT~\cite{bert} variants, have become the state-of-art solution for important machine learning tasks such as natural language processing~\cite{critical}, computer vision~\cite{image, objectdetection}, and video analysis~\cite{space}. These models have recently received special attention from the machine learning community for being extremely efficient in terms of both accuracy and performance. Transformers use attention mechanisms to gather information about the relevant context of a given input (i.e., a word of a sentence), and then encode that context in a vector. The attention mechanism allows to grab context information from distant parts of an input sequence to help understand its meaning, and it is implemented in the form of multiple feed-forward FC layers. However, the benefits of these networks come at the cost of long execution time due to the large memory footprint and low computation-to-memory access ratio. FC layers exhibit different characteristics with respect to CONV layers: weights are not reused by different neurons and the computation-to-memory access ratio is significantly smaller, i.e., FC layers are more memory intensive.

Each type of DNN is effective for a specific subset of cognitive applications. Moreover, for each application, each DNN has a different arrangement of layers with specific operations. FC and CONV layers take up the bulk of computations in most DNNs. Other types of layers performing pooling, normalization, or activation functions are also common. However, these other layers have no synaptic weights and represent a very low percentage of the DNN execution time. In this paper, we focus on optimizing the performance and energy efficiency of hardware accelerators for the inference of FC and CONV layers in modern MLPs, CNNs, RNNs, and Transformers.

\subsection{DNN Quantization}\label{DNN_QT}
Quantization is a highly popular technique to map values from a continuous range to a discrete set. The main purpose of quantization is to compress the original DNN models to reduce the memory footprint and the computational cost with a minor impact on accuracy. Equation~\ref{eqn:Quantization} shows an example of a function that quantizes real values (in floating-point, FP, precision) and maps them to an integer range.

\begin{equation}
\label{eqn:Quantization}
Q(r) = INT(r/s) - z 
\end{equation}

where \(Q(r)\) is the quantized value, \(r\) is a FP value, \(s\) is a scaling factor, and \(z\) is an integer offset. The \(INT\) function is a rounding to the nearest value. This method is also referred to as linear uniform quantization since the resulting quantized values (a.k.a. quantization levels) are uniformly spaced.

Recently, non-uniform quantization schemes have been proposed to further reduce the memory pressure. These methods have been designed for DNN models with tensors that have a bell-shaped long-tailed distribution of weights and activations~\cite{additive, long-tailed}. Logarithm quantization is en example of a non-uniform scheme, where the quantization levels increase exponentially instead of linearly~\cite{quantization_survay}. The Logarithmic Quantization (LQ)~\cite{L2L, convolutional, deep} offers smaller numerical precision (i.e. bitwidth) with lower accuracy loss compared to the linear quantization by exploiting the non-uniform distribution of tensors. QeiHaN employs uniform quantization for the weights, and a logarithmic base-2 (LOG2) quantization for the activations of all the FC/CONV layers. Section~\ref{analysis} provides more details on the LOG2 quantization and its main benefits.

\subsection{Dataflows in DNN Accelerators}
The dataflow of a DNN accelerator is defined as the mapping and scheduling of the computations as well as the data partitioning across compute units. The dataflow that is most effective to reduce the memory accesses and data movements to optimize performance and energy efficiency depends on the target cognitive computing task and hardware architecture~\cite{maeri}. The dataflow determines the storage requirements and communication patterns among main memory, local on-chip buffers inside PEs, and compute units.

In previous works~\cite{understanding_dataflow, eyeriss}, the election of the dataflow is based on minimizing the data movement of the inputs, outputs or weights. Therefore, DNN accelerators tend to follow one of these dataflows: Weight Stationary (WS), Output Stationary (OS), and Input Stationery (IS). In OS, each PE computes an output neuron at a time~\cite{shidiannao}. In the WS/IS dataflows, each PE pre-loads a set of weights/inputs from memory to local buffers, and those are used to perform all associated computations~\cite{scnn}.

QeiHaN uses an input stationary dataflow, which means that each input of a given layer is read and reused, until all the related computations are done, before reading the next input. The IS dataflow is the most suitable for our logarithmic quantization of DNN activations and efficient weight storage scheme. We compare QeiHaN with two baseline accelerators, one with OS dataflow and the other with IS dataflow.

\subsection{3D-Stacked Memory}
High-density 3D memory is a promising technology for the memory system of DNN and other domain-specific accelerators~\cite{application, guo3d}. It consists of stacking multiple memory dies on top of each other, which increases the memory capacity and bandwidth compared to 2D memory, and also reduces the access latency due to the shorter on-chip wiring interconnection. These aspects lead to an overall improvement in both energy efficiency and performance. The 3D memory dies are commonly based on DRAM, but the integration of other memory technologies is being actively researched with very promising results. On the other hand, recent advances in low-capacitance through-silicon vias (TSVs) technology have enabled 3D memory that includes a few DRAM dies on top of a logic chip, within a single package~\cite{tsv}. Although there are numerous implementations of 3D-stacked memory technologies, until now, the Hybrid Memory Cube (HMC)~\cite{HMC2_STD} by Micron and the High Bandwidth Memory (HBM)~\cite{HBM2, HBM} from AMD/Hynix are the preferred choices for most DNN accelerator proposals~\cite{neurocube, tetris, neurostream, transpim}.


HBM and HMC are designed for high performance data-centric applications. Both are composed of vertically stacked DRAM dies with a single logic layer at the bottom. These memory technologies take advantage of Through-Silicon Vias (TSVs) to enable high-bandwidth and low-latency communication between the stacked memory layers. In HBM, each DRAM die is partitioned horizontally, and different partitions on different dies are treated as independent memory channels. On the other hand, in HMC, each DRAM die is divided into multiple partitions in a 2D grid where the corresponding partitions in the vertical direction form a single vault. Both HBM and HMC can exploit memory-level parallelism by organizing the large number of TSVs into multiple independently-operated channels. This allows multiple partitions in the DRAM die to be accessed simultaneously, further enhancing memory bandwidth and overall system performance.

NDP systems employing HBM or HMC associate the PEs of the logic die with each channel or vault to efficiently utilize the memory-level parallelism and achieve high data processing throughput. The choice between HBM and HMC would depend on the specific requirements of the NDP system, and the desired trade-offs between memory bandwidth, energy efficiency, and integration with the host processor.


\subsection{3D-stacked DRAM-based DNN Accelerators}
\begin{figure}[t!]
\centering
\includegraphics[width=1.0\columnwidth]{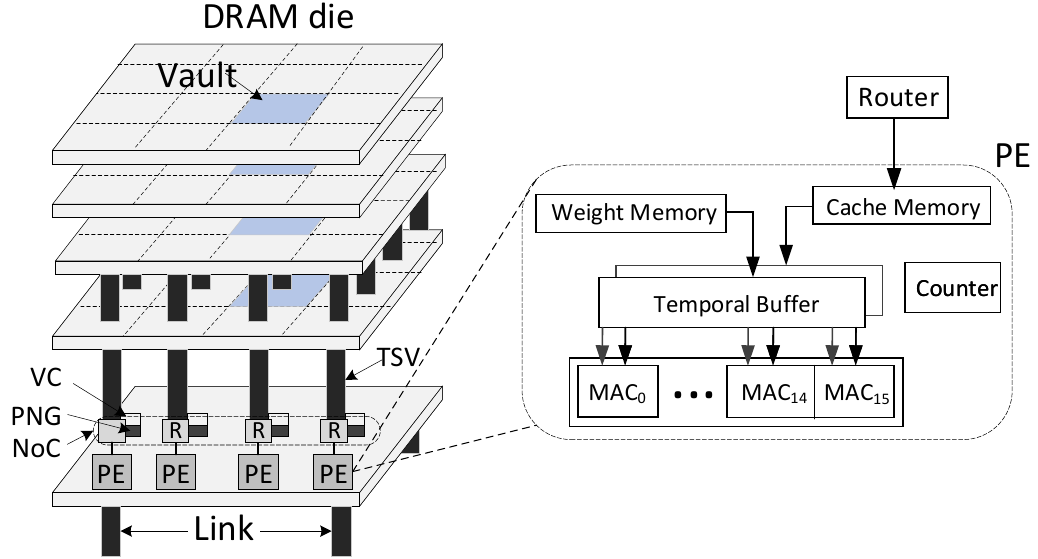}
\caption{Overview of the Neurocube architecture and organization of a PE (adapted from~\cite{neurocube}).}
\label{fig:neurocube}
\end{figure}

Neurocube~\cite{neurocube} is a programmable DNN accelerator integrated into the logic layer of a 3D stack DRAM-based HMC. The Neurocube architecture consists of clusters of processing engines (PE) connected by a 2D mesh NoC in the processing layer. Each PE of the logic layer is associated to a single memory vault, and can operate independently, and communicate through the TSVs and a vault controller (VC). The organization of each PE includes multiple memory buffers to store weights and inputs as well as some units to perform MAC operations. In addition, each vault controller includes a Programmable Neurosequence Generator (PNG) unit that generates the commands to orchestrate the corresponding operations of the DNN layers. The PNGs employ a simple finite state machine (FSM) with counters that are initialized depending on the number of MAC units in each PE and the DNN layer topology. Figure~\ref{fig:neurocube} shows a general overview of the Neurocube architecture and a PE. In this work, we implement a Neurocube-like baseline accelerator to assess the performance improvement and energy savings of QeiHaN.

In the same line of research, TETRIS~\cite{tetris} is another popular DNN accelerator based on HMC. Like Neurocube, TETRIS presents an optimized hardware architecture coupled with software scheduling and partitioning techniques that exploit the inherent characteristics of 3D memory. First, the authors show that the high throughput and low energy characteristics of 3D memory allow to rebalance the NN accelerator design, using more area for processing elements and less area for SRAM buffers. Second, they move some portions of the NN computations close to the DRAM banks to decrease the bandwidth pressure and increase performance and energy efficiency. Finally, they develop an optimized dataflow scheduler and hybrid partitioning scheme that parallelizes the DNN computations within and across multiple vaults and stacks.

%% file: sections/3-analysis.tex





\section{LOG2 Quantization Analysis}\label{analysis}
DNN quantization allows to reduce the numerical precision of activations and weights, which in turn favors the memory footprint and the computational cost of hardware accelerated DNN architectures. Therefore, quantization techniques have been widely explored in previous studies as described in Section~\ref{DNN_QT}. In particular, logarithmic quantization takes advantage of the non-uniform distribution of tensors to significantly reduce the numerical precision of input activations and/or weights with a minor impact in accuracy. This section analyzes the effects of the LOG2 quantization of activations on multiple DNN models and layers. First, we explore the benefit of the logarithmic encoding of activations to simplify the dot-product operations. Then, we provide some hints on reducing the number of accesses to the main memory by exploiting the characteristics of the 3D memory and bit-shifting operation.

Some prior works have used linear uniform quantization to compress the DNN parameters. However, we observe that activations and weights of most DNNs do not follow a uniform distribution, which causes a huge impact in terms of accuracy loss when the precision is further reduced to very low bitwidths (i.e. $<8b$). Specially in recent DNNs that are extremely deep and can have hundreds of layers, the error is propagated and expanded among layers.

On the other hand, logarithmic base-2 (LOG2) quantization~\cite{L2L, convolutional, deep} leverages the usually non-uniform distribution of activations and weights in a pre-trained DNN. The study in \cite{convolutional} compared the impact of linear and LOG2 quantization on activations and weights of VGG16 and AlexNet. Their analysis shows an exponential distribution of activation values around $0$. They also concluded that activations are more robust to LOG2 quantization than weights for several reasons. First, CONV layers reuse the weights multiple times when computing the dot-products, propagating the error across the inputs/outputs of all layers. Second, the range of the weights is not as wide as the activations~\cite{kernel_density}, and their density is often higher than that of the activations, that is, the amount of weights is huge, and their range is narrow. We performed an experiment applying LOG2 quantization to the activations and weights of modern DNNs, together and individually, and reached similar conclusions regarding weights being more sensitive to the LOG2 quantization error than activations. This suggests that the base-2 may not be the best fitting exponential base for quantizing the weights.

In this paper, we apply logarithmic base-2 (LOG2) quantization to the input activations of all the FC and CONV layers of a set of DNNs. On the other hand, we apply INT8 uniformly distributed linear quantization to the weights of these layers based on Equation~\ref{eqn:Quantization}. These layers represent close to 100\% of the total execution time for typical neural networks. Next, we analyze the distribution of exponents of the quantized activations, and the accuracy loss due to the LOG2 quantization. This scheme also allows us to efficiently re-organize offline the weights in-memory without additional expensive hardware, and exploit some of the intrinsic characteristics of the 3D-stacked memory, as described below. For each input $x$ and each layer $l$, the LOG2 quantization is applied according to the following equations:

\begin{equation}
\label{eqn:log_quant}
    LogQuant(x) =
    \begin{cases}
        0 & x= 0\\
        2^{\tilde{x}} & \text{otherwise}.
    \end{cases}
\end{equation}

\begin{equation}
\label{eqn:x_i}
\tilde{x} = Clip(Round(log_{2}(|x|))), min , max),
\end{equation}

where

\begin{equation}
\label{eqn:clip}
    Clip(x, min, max)= 
    \begin{cases}
        min & x\leqslant min\\
        max & x\geqslant max\\
        x & \text{otherwise}.
    \end{cases}
\end{equation}

The exponent $\tilde{x}$ is computed based on Equation~\ref{eqn:x_i}. The $Round$ function is defined as rounding to the nearest integer, and the clipping function in Equation~\ref{eqn:clip} forces the exponent values to be in the range of \([min, max]\), where $min = -(2^{n-1})$ and $max = (2^{n-1}-1)$. Assuming an n-bit exponential quantization (e.g. $n=4$), the number of unique intervals is $2^{n}-1$. We store an extra bit for the sign of the value, but in most layers it is not necessary since the activations are all positive. The $min$ exponent is also used as a special case to represent the exactly zero activation value, so all small activations are effectively pruned due to the clipping.

The main benefit of the LOG2 quantization is that it not only reduces the numerical precision but also eliminates the bulky digital multipliers by using simple shift and ADD operations. The approximated activation values $\tilde{x}$ are stored as exponents to reduce the memory pressure and the computational complexity. Equation~\ref{eqn:conv_in_log2} shows the transformed dot-product operation with the bit-shifting of $w_{i}$ weights by the $\tilde{x}_{i}$ exponents of the base-2 powers representing the activations, where ${x}_{i}$ is quantized to an integer exponent using Equation~\ref{eqn:log_quant}. Note that the positive exponents will lead to a shift to the left, while negative exponents result in a shift to the right.

\begin{equation}
\label{eqn:conv_in_log2}
\begin{split}
w^{T}x = \sum_{i=1}^{n} w_{i} \times x_{i} \simeq \sum_{i=1}^{n} w_{i}\times 2 ^ {\tilde{x}_{i}}\\
 = \sum_{i=1}^{n} Bitshift(w_{i}, \Tilde{x}_{i})
\end{split}
\end{equation}

In order to further exploit the LOG2 quantization of the input activations, a key observation is that, if a given activation is represented with a base-2 power of a negative exponent, the bit-shifting to the right will discard the least significant bits (LSB) of the weights that are multiplied by the corresponding activation. In other words, during the right bit-shift operation, and assuming that weights are uniformly quantized to 8 bits, only $1 \leqslant 8-|\tilde{x}| \leqslant 7$ bits of the weights are required while the rest can be avoided, reducing the memory accesses at a fine granularity.

\begin{figure}[t!]
\centering
\includegraphics[width=1.0\columnwidth]{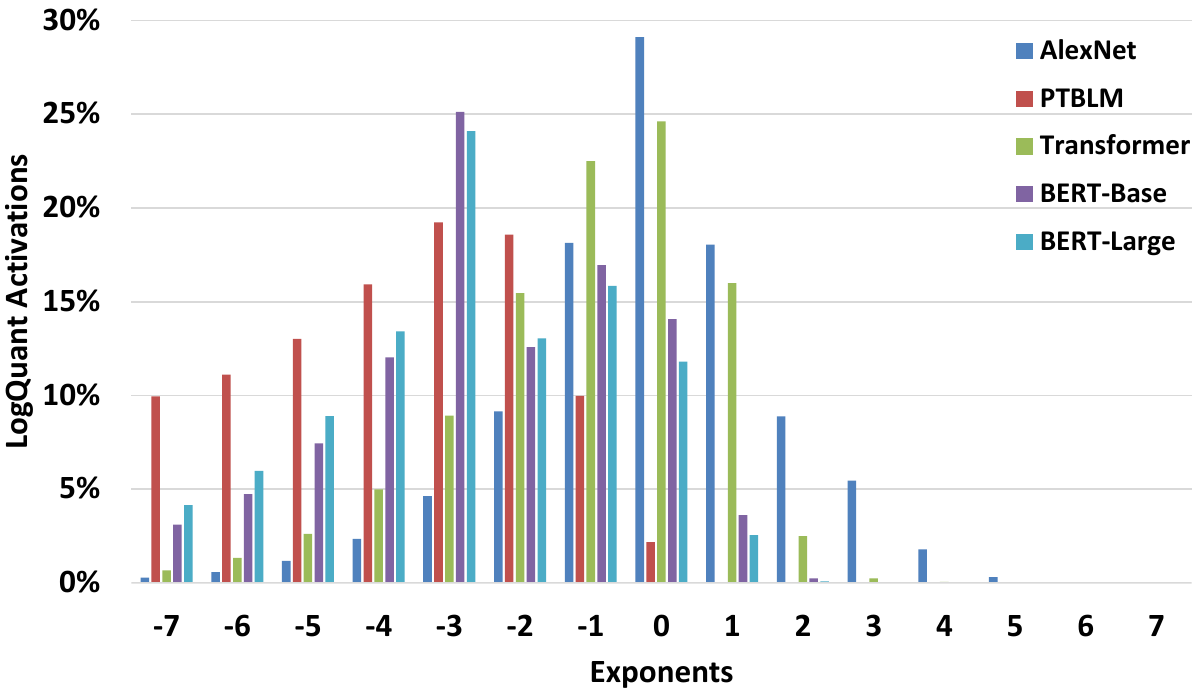}
\caption{Histograms of the LOG2 Quantization (LogQuant) of activations from all the FC and CONV layers of AlexNet, Transformer, PTBLM, BERT-Base and BERT-Large.}
\label{fig:activations}
\end{figure}

To demonstrate the potential of this idea, we perform an analysis of the exponents resulting from a LOG2 4-bit quantization of the activations in all FC/CONV layers of a popular set of DNNs from different domains. All the evaluated networks have been re-trained, reducing the accuracy loss after quantization to less than 1\% in all cases. Figure~\ref{fig:activations} shows the distribution of the non-zero, quantized activations. On average, more than 71\% of the activations have negative exponents. PTBLM (98\%), BERT-Base (82\%), and BERT-Large (85\%) have a similar distribution of exponents with a high concentration of negative values centered around $-3$, while the Transformer (57\%) and AlexNet (36\%) have the most symmetric distribution resulting in the lowest amount of negative exponents. QeiHaN, our proposed solution for efficient DNN inference, is based on exploiting this observation.



\begin{figure}[t!]
\centering
\includegraphics[width=1.0\columnwidth]{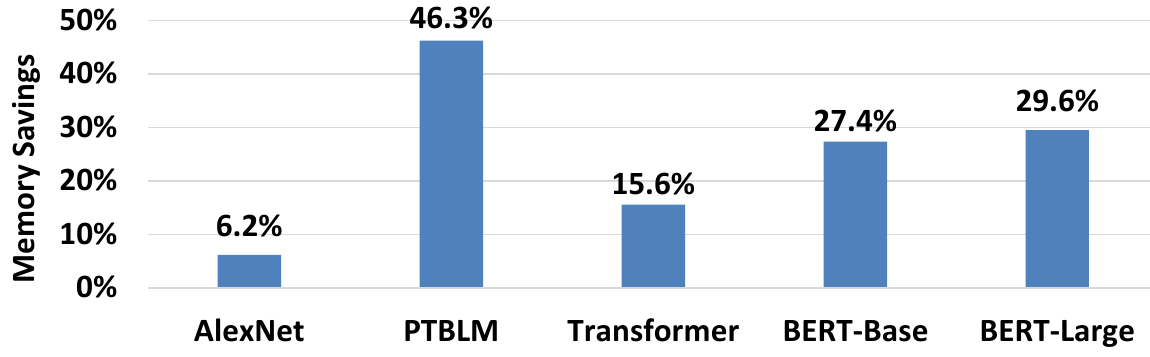}
\caption{Estimated memory savings for our set of DNNs.}
\label{fig:mem_saving}
\vskip -0.10in
\end{figure}

We define the \textbf{estimated memory savings} as the percentage of bits from the weights that can be ignored because the negative exponents of the base-2 activations render those bits useless when performing the bit-shifting operation. Figure~\ref{fig:mem_saving} shows that the memory savings are directly related to the histograms of the quantized activations. On average, 25\% of the memory accesses can be avoided. In addition, zero-activations are pruned in both, the baseline and our proposal, further reducing memory accesses. However, the conventional storage of weights in-DRAM is not suitable to exploit this optimization. The following section describes how to re-organize the weights in-memory to take full advantage of the LOG2 quantization.

%% file: sections/4-proposed_method.tex
\section{\NoCaseChange{QeiHaN} Accelerator}\label{QeiHaN Architecture}
This section describes the hardware support required to implement QeiHaN. First, we present the main hardware components of the QeiHaN accelerator. Next, we describe the memory organization of weights and activations. Finally, we show how FC and CONV layers are executed in the accelerator using QeiHaN with an enhanced input stationary dataflow.

\subsection{Architecture}\label{Architecture}
The goal of QeiHaN is to optimize the memory pressure by performing an implicit in-memory bit-shifting of the weights in the FC and CONV layers of different DNNs. QeiHaN leverages a large number of negative exponents after the LOG2 quantization of activations, and an efficient weight storage scheme, to save memory accesses. Similar to Neurocube~\cite{neurocube} and TETRIS~\cite{tetris}, QeiHaN is based on NDP architectures~\cite{new_DRAM_architecture, HMC2_STD} that leverage 3D stacked memory for high-performance, low-energy DNN inference. As described in Section~\ref{Preliminaries}, the 3D memory consists of multiple DRAM dies connected via TSVs to a logic die. DRAM dies are divided into vertical partitions named vaults that resemble conventional DDRx channels, which can operate independently. In addition, each vault is connected to a tile in the logic die to perform arithmetic computations on the stored data.


\begin{figure}[t!]
\centering
\includegraphics[width=1.0\columnwidth]{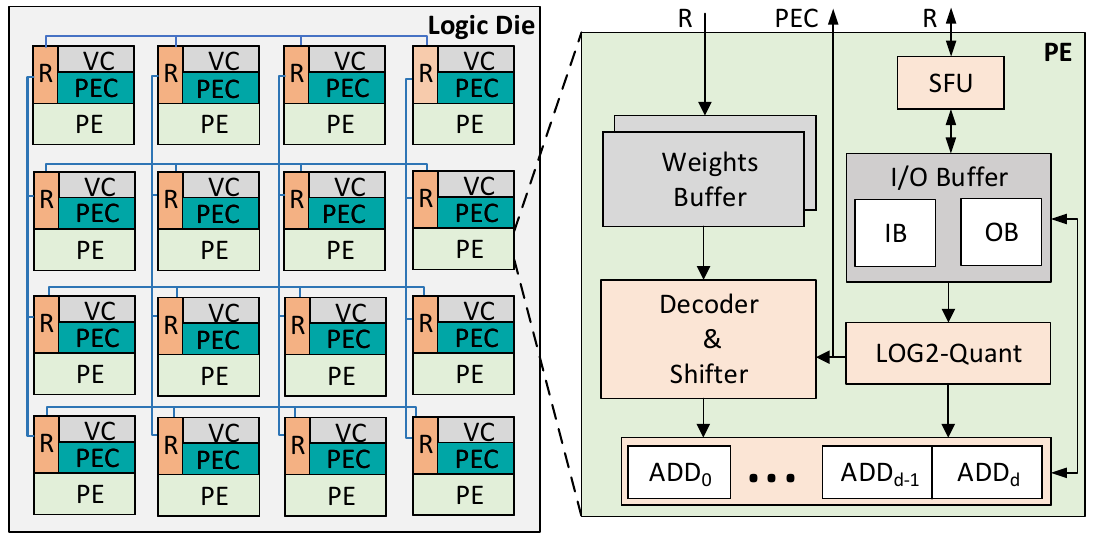}
\vskip -0.15in
\caption{Architecture of the QeiHaN accelerator including the organization of a single Processing Element (PE).}
\label{fig:QeiHaN}
\vskip -0.15in
\end{figure}

Figure~\ref{fig:QeiHaN} shows a high-level schematic of the QeiHaN architecture. Each tile in the logic die consists of a single PE, a Vault Controller (VC), a Router (R), and a PE Controller (PEC). The VC manages all the memory operations within the corresponding vault. The router provides local access between a given PE and its related vault, as well as remote access to the other vaults/PEs through a 2D mesh network. In addition, the PEC orchestrates the communication between the PE and the router by generating the addresses of the required data in each PE. Finally, the PE is the core of the tile, and is responsible for accelerating the DNN operations. The main components of a PE include the blocks of SRAM used for storing the inputs (IB), outputs (OB), and weights (WB), the LOG2 Quantization (LOG2-Quant) unit, the Weight Decoder and Shifter (D\&S) unit, the ADD array, and the Special Function Unit (SFU). Below is a detailed description of each component:

\textit{Memory Buffers:} Each PE in the logic die has three individual on-chip SRAM buffers to store and reuse the data fetched from the main memory according to the dataflow of the accelerator. First, a small Input Buffer (IB) stores blocks of input FP16 activations until filling the whole buffer space. Second, an Output Buffer (OB) stores the partial and final results that are produced during the execution of a DNN layer. Third, a Weights Buffer (WB) keeps the required bits of the weights for the bit-shifting operations. All the SRAM memories are double buffered to load data from main memory while performing computations, avoiding stalls in the pipeline, and highly multi-banked to achieve the bandwidth required to feed a large number of functional ADD units. In addition, all these buffers are sized considering the worst case scenarios, that is, the biggest layer for the I/O buffer, and all the 8 bits of $M$ weights for the WB, where $M$ is the bus size of a vault in the 3D-stacked memory.

\begin{figure}[t!]
\centering
\includegraphics[width=0.8\columnwidth]{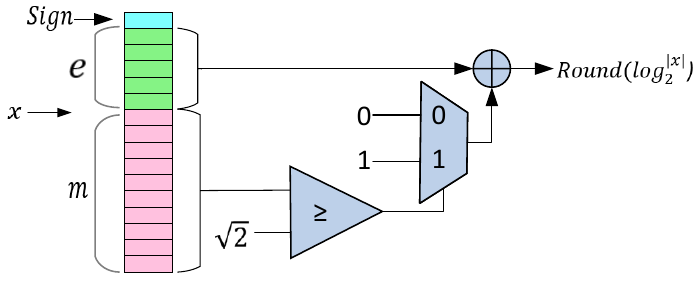}
\caption{Hardware implementation of $Round(log_{2}|x|)$.}
\label{fig:log2_imp}
\end{figure}

\textit{LOG2-Quant Unit:} This unit is in charge of the LOG2 quantization of input activations from FP16 to base-2 exponential values according to Equation~\ref{eqn:log_quant}. Figure~\ref{fig:log2_imp} shows the hardware required to compute the $Round(log_{2}(|x|))$ function of Equation~\ref{eqn:x_i}. Unlike previous works that use relatively complex hardware~\cite{log2_im1,log2_im2, ip}, we implement this function with a very simple scheme. In particular, we perform a comparison between the fractional part of the value $|x|$ and the $\sqrt{2}$ using a simple comparator. The standard half precision (FP16) format of a value $x$ is encoded with a sign bit, mantissa $m$, and exponent $e$. The exponent $e$ is already expressed as an integer in base-2 format, so the LOG2 function of $|x|$ can be implemented by applying the logarithm on the mantissa $m$ as shown in Equation~\ref{eqn:log_comp1}. Taking into account the hidden bit of the mantissa, $m$ is always a value between [1, 2). Therefore, the term $Round(log_{2}m)$ can be further simplified by Equation~\ref{eqn:log_comp2}. In the next step, each quantized value $\tilde{x}$ in QeiHaN is represented by a 4-bit exponent through a clipping function (i.e. Equation~\ref{eqn:clip}), resulting in a range of $[-8, 7]$. An extra bit may be used for the actual sign of the activations, except for when all are known to be positive due to the ReLU activation function. In addition, all zero activations will skip the quantization and all the related computations and memory accesses. Similarly, all the small activations clipped to $-8$ will be effectively pruned (rounded to zero). Finally, each quantized activation is sent to the D\&S unit for further processing.

\begin{equation}
\label{eqn:log_comp1}
Round(log_{2} |x|) = e + Round(log_{2} m)
\end{equation}

\begin{equation}
\begin{footnotesize}
\label{eqn:log_comp2}
1 \leqslant m < 2 \xrightarrow{} 0 \leqslant log_{2} m < 1 \Rightarrow Round(log_{2}m) =
\begin{cases}
0 & m < \sqrt{2}\\
1 & m \geqslant \sqrt{2}
\end{cases}
\end{footnotesize}
\end{equation}

\textit{Weight Decoder \& Shifter Unit (D\&S):} The weights that multiply non-zero activations are decoded from a compressed stream and bit-shifted by appending the necessary amount of zeros based on the exponent from the LOG2-Quant unit. According to the exponent value $\tilde{x}$, the PE controller determines the required bits of the weights that have to be loaded from DRAM and stored into the WB. A non-negative exponent requires loading all 8 bits of each weight, and the D\&S unit shifts the weights $\tilde{x}$ positions to the left before sending the results to the ADD array. Otherwise, we only need to fetch the $8-|\tilde{x}|$ MSBs of the weights. For example, given a negative exponent $\tilde{x}=-3$, only the 5 MSBs of each weight are loaded into the WB. Then, the D\&S unit reads the selected bits of the weights from the WB and generates a set of 16-bit $d$ values, where $d$ is the amount of adders in the ADD array. Note that the bit-shifted weights are the result of the traditional multiplication of activations and weights. In order to use this unit efficiently, QeiHaN reorganizes the weights in-memory to a bit-level granularity as described below in Section~\ref{mem_org}.

\textit{ADD Array:} This array is made of $d$ independent ADD units that are used to accumulate the products of each activation by the corresponding weights. According to the sign of the activation value, not the exponent, the bit-shifted weight is added/subtracted to/from the partial outputs computed in previous cycles and stored in the OB. The LOG2 quantization removes the need for any multiplier, so the partial outputs are loaded from the OB and the bit-shifted weights come from the \textit{D\&S}. As a result, in a single execution all the adders compute partial outputs related to the same input activation from $d$ different convolutional kernels or output neurons.

\begin{figure}[t!]
\centering
\includegraphics[width=0.85\columnwidth]{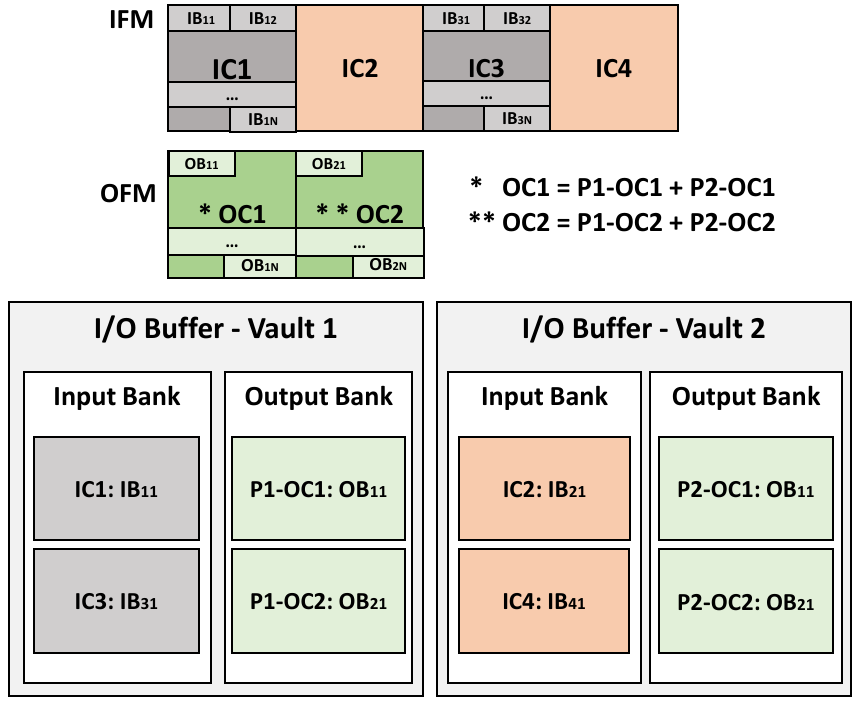}
\caption{Organization of the I/O Buffer.}
\label{fig:I/O_Buffer_Organization}
\end{figure}

\textit{Special Function Unit (SFU):} The SFU is composed of units to perform non-linear activation functions, pooling, and normalization, among others. These functions are usually applied to the final outputs of the FC/CONV layers at the end of their execution, and tend to require more numerical precision in order not to lose accuracy. Thus, QeiHaN de-quantizes the resulting 16-bit integer outputs back to FP16 before using those functions. The non-linear functions are implemented with Look-Up-Tables (LUTs).

\subsection{Memory Organization}\label{mem_org}
This section describes the memory organization of the accelerator, which refers to the data layout of weights and activations inside the DRAM of each vault and the on-chip buffers of the PEs.

To illustrate it, the top of Figure~\ref{fig:I/O_Buffer_Organization} shows an example of a small CONV layer with an input feature map (IFM) size of four channels (IC1-IC4), and an output feature map (OFM) size of two channels (OC1-OC2). On the other hand, the bottom of Figure~\ref{fig:I/O_Buffer_Organization} shows how the input/output activations of the different channels are partitioned and distributed among the I/O buffers of two different PEs/Vaults.

In QeiHaN, the input activations are divided channel-wise across all vaults, that is, all inputs of a given channel are stored in the same vault. In contrast, each vault allocates a portion of the corresponding partial outputs of all the channels. In CONV layers, the dimensionality of the inputs/outputs may be quite large, so we employ a blocking scheme to reduce the on-chip storage requirements by segmenting the IFM and OFM into $N$ blocks or tiles per channel. The I/O Buffer only stores a subset of blocks for each assigned channel of the IFM and OFM, the block size being significantly smaller than the dimensions of the feature maps. Note that each Vault/PE is working on a different set of inputs but producing partial outputs of the same OFM channels. Hence, a reduction is required at the end of the execution to obtain the final outputs. Likewise, FC layers are a special case of CONV, where there is just a single block and input per channel (i.e. $N=1$).

\begin{figure}[t!]
\centering
\includegraphics[width=1.0\columnwidth]{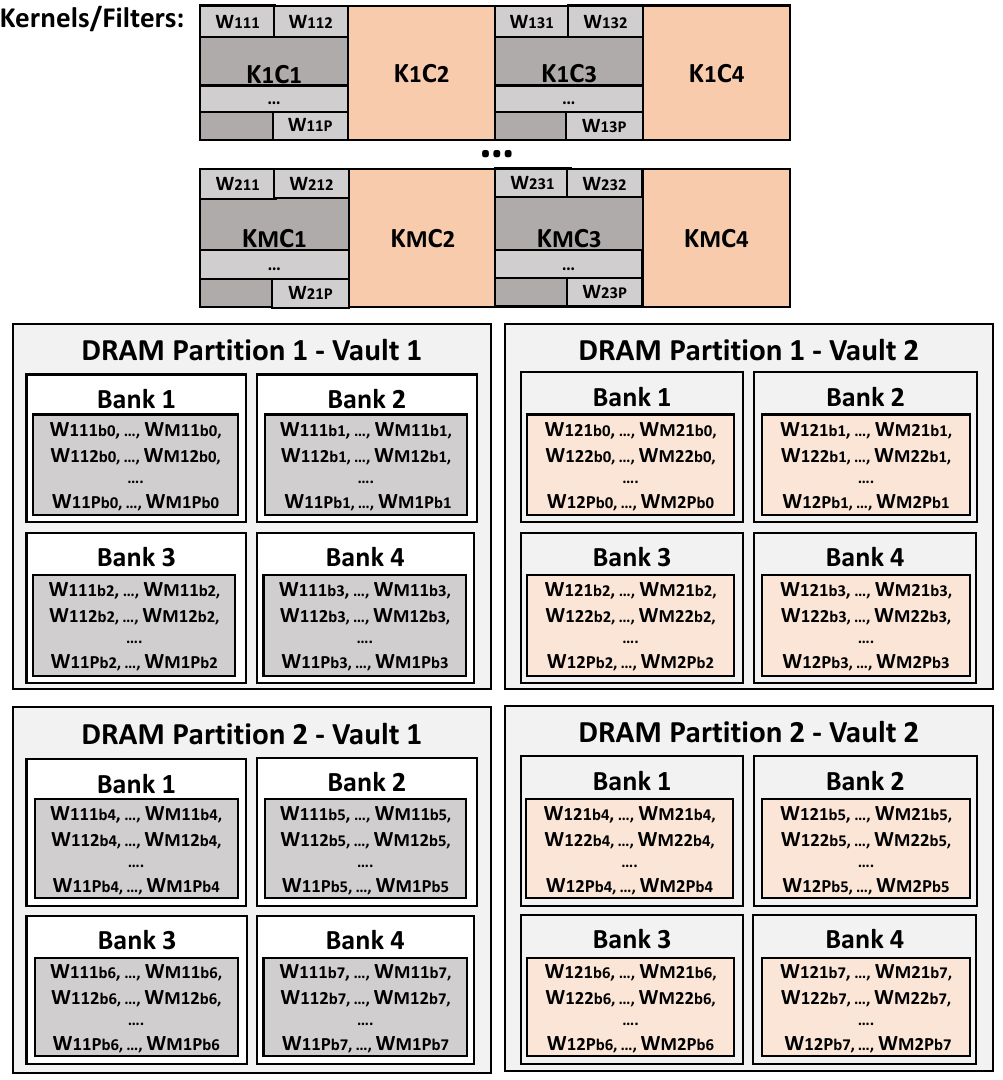}
\caption{Organization of weights in the DRAM-based 3D-stacked memory.}
\label{fig: Weight Organization}
\end{figure}

Figure~\ref{fig: Weight Organization} shows the layout of $M$ filters or kernels with $P$ weights per channel each in the DRAM dies of each vault, where each partition includes 4 banks, for the example of Figure~\ref{fig:I/O_Buffer_Organization}. Similar to the activations, the weights of each kernel are also distributed channel-wise across all vaults. The bits of the weights of the corresponding channels are interleaved in the different banks and partitions of the same vault. That is, the least-significant bit (LSB) of a subset of weights is stored in the first bank of a vault, then the next bit in the second bank and so on. This layout simplifies the implementation of our implicit bit-shifting scheme, as it is easy to locate all the bits of the weights that are required to operate with a given input in case some have to be skipped and others accessed.

In addition, most 3D-stacked DRAM-based operations use a \textbf{Closed-Page Policy} to reduce power consumption~\cite{demystifying}. Consequently, applications benefit from \textbf{Bank-Level Parallelism} but not from spatial locality. QeiHaN remaps the data to avoid internal organization bottlenecks and, hence, requests to different banks can be concatenated/overlapped to effectively achieve high bandwidth. Note that weights are known statically so their organization can be pre-arranged offline.

\begin{figure*}[t!]
\centering
\includegraphics[width=0.75\textwidth]{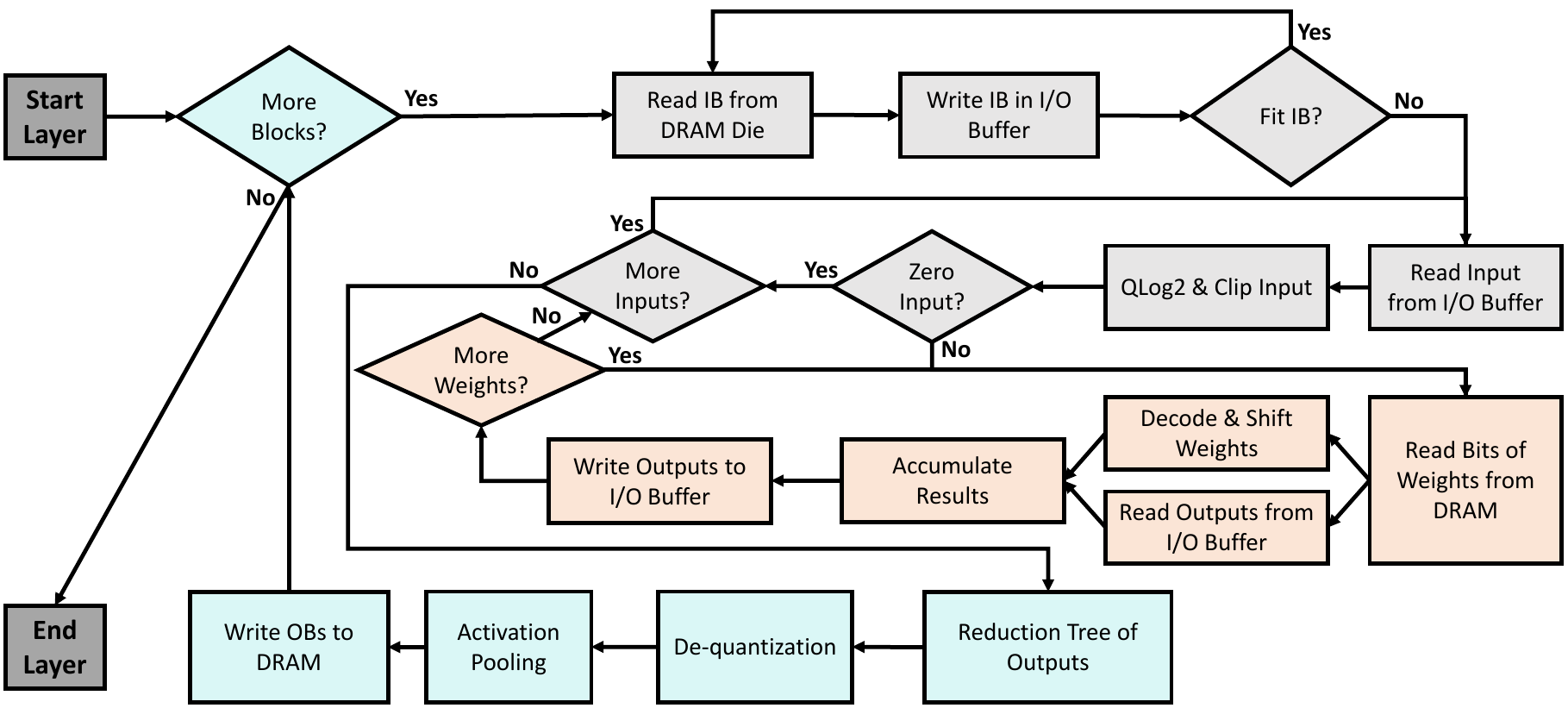}
\vskip -0.10in
\caption{Dataflow and Execution scheme.}
\label{fig:flowchart}
\vskip -0.15in
\end{figure*}

\subsection{Dataflow}
Neurocube~\cite{neurocube} follows an output stationary (OS) dataflow in which each PE computes a subset of outputs at a time. This dataflow is inefficient to exploit the resources of the 3D memory, as demonstrated by our results in Section~\ref{evaluation}. On the other hand, QeiHaN uses an enhanced input stationary (IS) dataflow coupled with a blocking scheme to efficiently exploit the LOG2 quantization of the input activations, minimizing the memory accesses to both weights and activations. Figure~\ref{fig:flowchart} illustrates the dataflow of the QeiHaN accelerator with a flowchart. The proposed dataflow includes three main stages marked in different colors: Pre-Processing (Gray), Execution (Orange), and Post-Processing (Blue).

In the \emph{Pre-Processing} stage, each PE reads input activations from DRAM until filling the input buffer space. That is, inputs (outputs) are pre-loaded (processed) on-demand by blocks, activations are stored in FP16 format, and the size of the blocks is computed according to the feature map sizes and the I/O buffer capacity. In the IS dataflow, each PE of the accelerator fetches and processes one input of a block at a time from the I/O buffer, and performs all the associated computations before moving to the next input. First, the LOG2 quantization and clipping function is applied to obtain the 4-bit exponent $\tilde{x}$. Then, QeiHaN also performs a zero and small activation pruning. Concurrently, the reading of input blocks from DRAM continues in the background, as long as there is space in the buffers, to hide the memory latency while doing computations of the current blocks.

In the \emph{Execution} stage, and based on the value of the exponent $\tilde{x}$, a set of $M$ useful bits of INT8 uniformly quantized weights of $M$ different kernels related to the input are read from DRAM at a time, where $M$ is determined by the internal 3D-stacked memory bus size (e.g. 32-bit). Thus, in each request, the bits in the same position for $M$ different weights are loaded into the weights buffer, and multiple requests are made until all the required bits are retrieved. Next, the bits of the weights are decoded and bit-shifted by appending the corresponding zeros, resulting in 16-bit integer values. These results are grouped and sent to the ADD array unit in batches of $d$ values, where $d$ is the number of adders (e.g. 16). In parallel, the partial outputs from previous executions are loaded from the output buffer. Then, the accelerator performs $d$ ADDs to accumulate the results of each output with the shifted weights, followed by the write-back to the output buffer. This stage is repeated until all the weights of all filters related to the current activation are processed.

Finally, in the \emph{Post-Processing} stage, QeiHaN reduces the partial outputs of each PE. The reduction starts as soon as enough activations complete all their operations. Then, in a centralized PE, the final results are de-quantized, and the SFU performs the activation and pooling operations before distributing and storing the corresponding activations back to each vault. After processing all the blocks of inputs the layer execution is completed. Note that all the main steps are carried out in parallel in a deep pipeline.

%% file: sections/5-methodology.tex
\section{Methodology}\label{methodology}
This section presents the methodology for evaluating QeiHaN, our NDP accelerator for DNN inference.

\textbf{Workloads.} Our objective is to prove that our scheme provides important savings for multiple applications and different DNN models. To this end, we evaluate QeiHaN on five state-of-the-art DNN workloads from different domains, summarized in Table~\ref{t:models}. Their model sizes range from medium to large scale with several hundreds of MBytes in memory footprint. In particular, we include the ILSVRC 2012 winner, AlexNet~\cite{alexnet} (5 CONV and 3 FC layers), one of the most popular CNNs for image classification with the ImageNet dataset, and PTBLM~\cite{ptblm} (2 LSTM layers), an RNN that consists of LSTM cells for language modeling using the Penn Treebank dataset. In addition, we employ three attention-based networks: Transformer (6 Encoders, 6 Decoders), BERT-Base (12 Encoders, 110M Parameters), and BERT-Large (24 Encoders, 340M parameters). The Transformer~\cite{attention} model is evaluated on the machine translation task of Newtest2014 (English to German) which contains 3003 sentences. BERT-Base~\cite{bert}, and its larger variant BERT-Large, are evaluated on the question-answering task of SQuADv1~\cite{squad}. Finally, all these networks have been re-trained in order to recover the accuracy after quantization, that is, less than 1\% loss. Accuracy is reported as Top-1 for image classification (higher is better), perplexity for language modeling (lower is better), bilingual evaluation understudy (BLEU) for machine translation (higher is better), and weighted average of the precision and recall (F1) for question-answering (higher is better).


\begin{table}[t!]
\caption{DNNs employed for the experimental evaluation of QeiHaN. The model size accounts for the parameters in FP32 and INT8 of the FC and CONV layers where the LOG2 quantization is applied. The accuracy shown is after quantization.}
\label{t:models}
\vskip -0.20in
\begin{center}
\resizebox{1.0\columnwidth}{!}{%
    \begin{tabular}{cccc}
    \hline
    \textbf{DNN Model}              &    \textbf{FP32-Size (MB)}    &    \textbf{INT8-Size (MB)}       &      \textbf{Accuracy}      \\
    \hline
    AlexNet~\cite{alexnet}          &                 144          &                 36                &        57.05\% (Top-1)      \\
    PTBLM~\cite{ptblm}              &                 136          &                 34.2              &        9.40 (Perplexity)    \\
    Transformer~\cite{attention}    &                 336          &                 84                &        28.61 (BLEU)         \\
    BERT-Base~\cite{bert}           &                 440          &                 110               &        86.75\% (F1)         \\
    BERT-Large~\cite{bert}          &                 1320         &                 330               &        89.72\% (F1)         \\
    \hline
    \end{tabular}%
}
\end{center}
\vskip -0.20in
\end{table}

\textbf{System models and simulation.} We have developed a simulator that accurately models three different systems, QeiHaN and two baseline accelerators. The first baseline is inspired in Neurocube~\cite{neurocube}, described in Section~\ref{Preliminaries}, but with some optimizations, such as a lower quantization bitwidth, to isolate the effects of our proposal when comparing the two. The second baseline, named NaHiD, implements the same architecture, dataflow, and quantization scheme as QeiHaN but with a standard memory organization of the weights. That is, NaHiD also replaces multiplications by bit-shift operations and additions but, in contrast to QeiHaN, it requires loading all the bits of the weights from memory. This comparison allows us to infer the main benefits due to the QeiHaN's efficient 3D memory-centric weight storage scheme. Table~\ref{t:parameters} shows the parameters of the experiments. For a fair comparison, we set most of the configuration parameters to match the Neurocube baseline: a 3D-stacked memory of 4 GB with 4 DRAM dies partitioned into $4 \times 4$ vaults and PEs, an internal 3D memory bandwidth of 10 GB/s per vault, about 2.5 KB of SRAM per PE, 16 MAC/ADD units per PE, and a frequency of 300 MHz in the logic die. QeiHaN and NaHiD require slightly smaller memory buffers (i.e. 2KB of OB, 64B of IB, and 64B of WB) due to the different dataflow.

\begin{table}[t!]
\caption{Parameters for the accelerators.}
\label{t:parameters}
\begin{center}
\resizebox{0.75\columnwidth}{!}{%
    \begin{tabular}{cc}
    \hline
    \multicolumn{2}{c}{\textbf{Common Parameters}}                     \\
    \hline
    Technology                                      &  32 nm            \\
    Logic Die Frequency                             &  300 MHz          \\
    \#DRAM Dies, \#Banks per Vault per Die          &  4                \\
    \#Vaults, \#PEs                                 &  16               \\
    3D-stacked Memory Total Size                    &  4 GB             \\
    3D-stacked Memory Bandwidth per Vault           &  10 GB/s          \\
    \hline
    \multicolumn{2}{c}{\textbf{Neurocube Parameters}}             \\
    \hline
    Dataflow                                        &  OS               \\
    Weights Precision                               &  8-bit            \\
    Input Activations Precision                     &  8-bit            \\
    Weights/Inputs Quantization                     &  Uniform/Uniform  \\
    \#MAC Units per PE                              &  16               \\
    Total SRAM Buffers Size per PE                  &  2.5 KB           \\
    \hline
    \multicolumn{2}{c}{\textbf{NaHiD \& QeiHaN Parameters}}     \\
    \hline
    Dataflow                                        &  IS               \\
    Weights Precision                               &  8-bit            \\
    Input activations Precision                     &  4-bit            \\
    Weights/Inputs Quantization                     &  Uniform/LOG2     \\
    \#ADD units per PE                              &  16               \\
    Total SRAM Buffers Size per PE                  &  2.1 KB           \\
    \hline
    \end{tabular}%
}
\end{center}
\end{table}

Regarding area and energy consumption evaluation, the logic components are implemented in Verilog, including all the additional components required by QeiHaN, and synthesized to obtain the delay, area, and power using the Synopsys Design Compiler~\cite{Designcompiler}, the modules of the DesignWare library and the technology library of 28/32nm from Synopsys. On the other hand, we characterize the memory buffers of the accelerator by obtaining the delay, energy per access, and area using CACTI-P~\cite{cacti}. We use the configurations optimized for low power and a supply voltage of 0.78V. Finally, the energy consumption of the 3D-stacked memory is estimated by using an HMC configuration of DRAMSim3~\cite{dramsim3}. The results obtained with the aforementioned tools are combined with the activity factors and memory traces provided by our simulator to obtain the dynamic and static power of the accelerators.

%% file: sections/6-evaluation.tex








\section{Evaluation}\label{evaluation}
This section evaluates the performance, energy efficiency, and memory activity of our proposal. First, we introduce an analysis of the total number of memory accesses to the 3D-stacked DRAM dies after applying the QeiHaN scheme. Then, we present the speedups and energy savings achieved by QeiHaN compared to the \textit{Neurocube} and \textit{NaHiD} baselines. Finally, we discuss the accelerator overheads.

\subsection{3D-stacked Memory Accesses}
\begin{figure}[t!]
\centering
\includegraphics[width=1.0\columnwidth]{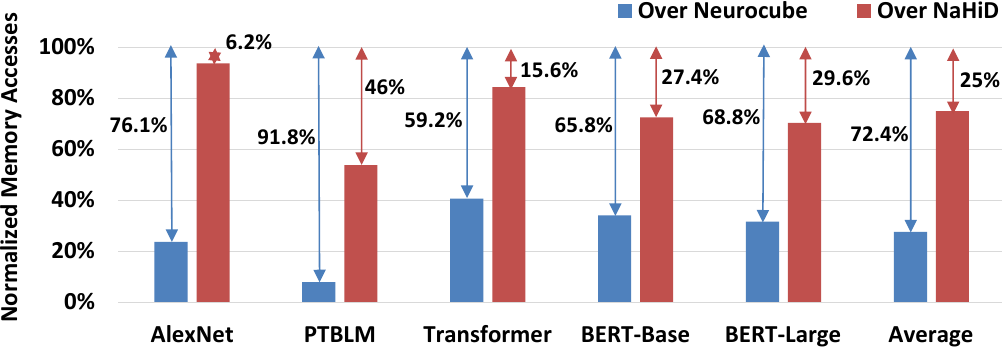}
\caption{Normalized total memory accesses for each DNN.}
\label{fig:total_hmc_accesses}
\end{figure}

Figure~\ref{fig:total_hmc_accesses} reports the normalized total 3D memory accesses of QeiHaN over the two baseline accelerators. This total includes both, memory accesses for reading/writing the weights and the input activations. On average for our set of DNNs, QeiHaN reduces the total DRAM accesses by 72.4\% and 25\% over \textit{Neurocube} and \textit{NaHiD}, respectively. The great reduction of memory accesses with respect to the baselines is mainly due to constraining the accesses to only the required bits of the weights for the bit-shifting operations. Moreover, QeiHaN shows a higher reduction of memory accesses over \textit{Neurocube} due to two main reasons. First, the enhanced IS dataflow of QeiHaN requires each input activation to be accessed just once during the execution of a layer. In contrast, the OS dataflow of \textit{Neurocube} may require multiple accesses to the activations. Second, QeiHaN performs pruning of zero and small activations after applying the quantization, removing all the related memory accesses to the weights. The efficiency of the activation pruning is limited in \textit{Neurocube} due to its OS dataflow, so it is not implemented. On the other hand, compared to \textit{NaHiD}, the reduction is well correlated to the estimated memory savings due to the huge amount of negative exponents as discussed in Section~\ref{analysis}. Both QeiHaN and \textit{NaHiD} use the same dataflow and pruning scheme, so both access the same input activations, and the savings come from the weights.

\subsection{Performance}
\begin{figure}[t!]
\centering
\includegraphics[width=1.0\columnwidth]{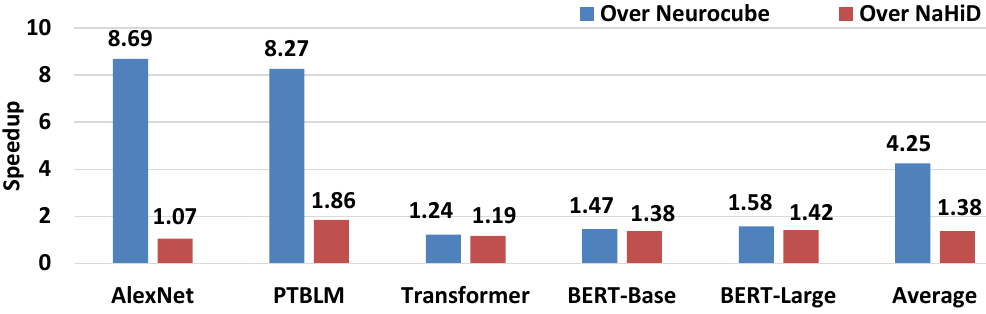}
\caption{Speedups of QeiHaN over \textit{Neurocube} \& \textit{NaHiD}.}
\label{fig:speedup}
\end{figure}

Figure~\ref{fig:speedup} shows the speedups achieved by QeiHaN. Compared to \textit{Neurocube}, QeiHaN provides consistent speedups for the five DNNs that range from $8.69x$ (\textit{AlexNet}) to $1.24x$ (\textit{Transformer}), achieving an average performance improvement of $4.25x$. The reduction in execution time is due to QeiHaN's efficient memory organization and enhanced IS dataflow. The number of memory accesses is dramatically reduced since only the meaningful bits of the weights required by the shift operations are loaded. In addition, QeiHaN employs a novel weight storage scheme to exploit the bank-level parallelism of the 3D memory. Moreover, QeiHaN overlaps the different stages of the dataflow in a deep pipeline, shortening the critical path of the execution. As shown in Figure~\ref{fig:total_hmc_accesses}, \textit{AlexNet} and \textit{PTLBM} exhibit the highest reduction in memory accesses and, hence, they obtain the largest performance improvements. The difference in speedup between these two networks and the attention-based models is in the percentage of zero and small activations that are effectively pruned in QeiHaN, skipping part of the execution and post-processing stages. The effect of activation pruning is minor in \textit{Transformer} (3\%), \textit{BERT-Base} (7\%), and \textit{BERT-Large} (13\%), but significant in \textit{AlexNet} (47\%) and \textit{PTLBM} (55\%).

Compared to \textit{NaHiD}, the benefits of QeiHaN are more modest but still quite important, achieving an average speedup of $1.38x$. The main reason is that both accelerators benefit from the same architecture, dataflow, quantization, and activation pruning scheme. Therefore, the improvements come mainly from the novel memory layout for storing the weights in the 3D memory, and the corresponding reduction of memory accesses by leveraging the logarithmic quantization. \textit{PTBLM} obtains the largest benefits, achieving an speedup of $1.86x$ whereas \textit{AlexNet} gets the lowest improvements, that is, $1.07x$ speedup. These results are directly proportional to the percentage of negative exponents shown in Figure~\ref{fig:activations}.

\subsection{Energy Consumption}
\begin{figure}[t!]
\centering
\includegraphics[width=1.0\columnwidth]{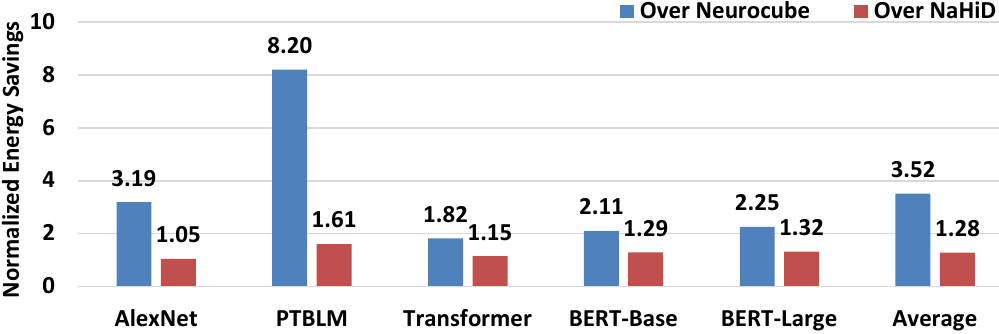}
\caption{Normalized energy savings for each DNN.}
\label{fig:energy}
\end{figure}

Figure~\ref{fig:energy} reports normalized energy savings. On average, QeiHaN reduces the energy consumption of the accelerator by $3.52x$ and $1.28x$ over \textit{Neurocube} and \textit{NaHiD}, respectively. As we observed for performance, the energy savings are well correlated with the number of negative exponents and the corresponding reduction of memory accesses. These energy savings are due to two main reasons. First, dynamic energy is reduced due to the savings in multiplications and memory accesses. Second, the performance improvements shown in Figure~\ref{fig:speedup} provide a reduction in static energy. Again, \textit{PTBLM} obtains the largest benefits, achieving a reduction of $8.2x$ and $1.6x$ in energy compared to both \textit{Neurocube} and \textit{NaHiD} respectively.

\begin{figure}[t!]
\centering
\includegraphics[width=1.0\columnwidth]{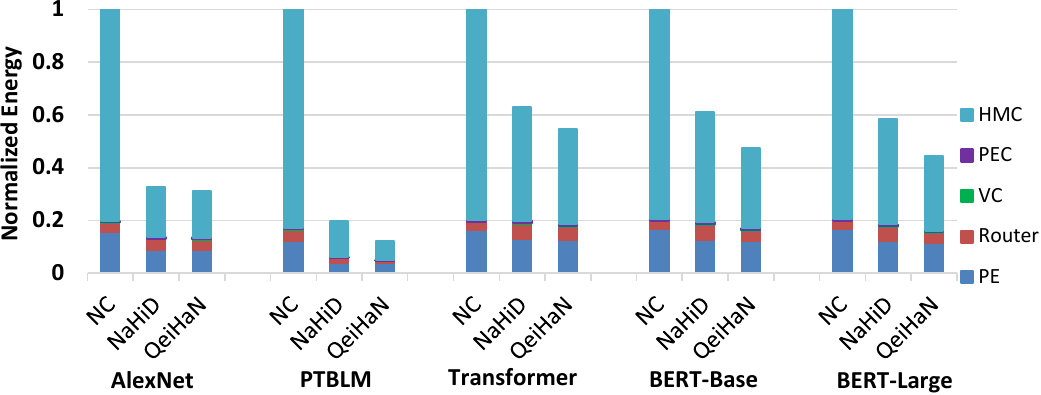}
\caption{Energy breakdown of QeiHaN and \textit{NaHiD} normalized to the \textit{Neurocube (NC)} baseline.}
\label{fig:energy_breakdown}
\end{figure}

Figure~\ref{fig:energy_breakdown} shows the energy breakdown of QeiHaN and \textit{NaHiD} over \textit{Neurocube}. The figure shows results for the five neural networks including the percentage of energy consumed by each major hardware block of the accelerators. As can be seen, the DRAM of the 3D-stacked memory (i.e. HMC) consumes most of the energy in all cases. The energy savings achieved by our proposal are significant, and are especially large in the 3D memory, since our scheme provides important savings in memory accesses for fetching the synaptic weights. In addition, the replacement of multipliers by simple bit-shift logic also results in smaller energy in the PEs. Note that the energy required for performing the logarithmic quantization is also included in the energy consumption of the PEs.

\subsection{Area}\label{subs:overheads}
QeiHaN requires extra hardware in the PEs of the accelerator to perform the LOG2 quantization of activations. As shown in Figure~\ref{fig:log2_imp}, we implement the \textit{LOG2-Quant} unit with a single comparator, one multiplexer, and one integer adder. These units represent less than 0.1\% of the total area and energy. In addition, we replace the costly multipliers by simple bit-shift logic, and the size of the SRAM buffers is also smaller, reducing the computational cost and the overall area of the PEs. The area overhead of QeiHaN in the logic die due to 16 PEs is $0.389mm^2 (16 \times 0.024mm^2)$ in $32nm$. We can see that QeiHaN with 16 PEs fits in a small part of the logic die ($68mm^2$~\cite{new_DRAM_architecture}) of the 3D stack. In comparison, \textit{Neurocube} extra storage and multipliers result in 20\% more area than QeiHaN at the same technology node (i.e. $0.487mm^2$). We do not evaluate the thermal constraints of QeiHaN since we expect them to be similar or lower than \textit{Neurocube} due to the smaller area of our accelerator.

%% file: sections/7-conclusion.tex
\section{Conclusions}\label{conclusion}
In this paper, we show that the distribution of activations among different FC and CONV layers of a representative set of modern DNNs exhibits a high degree of negative exponents after the logarithmic quantization, resulting in a high number of right bit-shift operations. Then, we propose QeiHaN, a new 3D-stacked DRAM-based NDP accelerator that exploits the log quantization to replace multiplications and reduce memory accesses to only the useful bits of the weights. QeiHaN implements an implicit in-memory bit-shifting of the DNN weights coupled with an efficient weight storage scheme. We show that QeiHaN requires minor hardware changes over Neurocube, a state-of-the-art accelerator, mainly an additional quantization unit made of a small set of comparators. Our experimental results show that, on average, QeiHaN provides $3.5x$ energy savings and $4.3x$ speedup with negligible accuracy loss and lower area than Neurocube.

%% file: sections/8-acknowledgement.tex
\section{Acknowledgement}\label{acknowledgement}
This work has been supported by the CoCoUnit ERC Advanced Grant of the EU’s Horizon 2020 program (grant No 833057), the Spanish State Research Agency (MCIN/AEI) under grant PID2020-113172RB-I00, and the ICREA Academia program.

%% file: main.bbl
\begin{thebibliography}{10}
\providecommand{\url}[1]{#1}
\csname url@samestyle\endcsname
\providecommand{\newblock}{\relax}
\providecommand{\bibinfo}[2]{#2}
\providecommand{\BIBentrySTDinterwordspacing}{\spaceskip=0pt\relax}
\providecommand{\BIBentryALTinterwordstretchfactor}{4}
\providecommand{\BIBentryALTinterwordspacing}{\spaceskip=\fontdimen2\font plus
\BIBentryALTinterwordstretchfactor\fontdimen3\font minus
  \fontdimen4\font\relax}
\providecommand{\BIBforeignlanguage}[2]{{%
\expandafter\ifx\csname l@#1\endcsname\relax
\typeout{** WARNING: IEEEtranS.bst: No hyphenation pattern has been}%
\typeout{** loaded for the language `#1'. Using the pattern for}%
\typeout{** the default language instead.}%
\else
\language=\csname l@#1\endcsname
\fi
#2}}
\providecommand{\BIBdecl}{\relax}
\BIBdecl

\bibitem{aga}
S.~Aga, N.~Jayasena, and M.~Ignatowski, ``Co-ml: a case for co llaborative ml
  acceleration using near-data processing,'' in \emph{Proceedings of the
  International Symposium on Memory Systems}, 2019, pp. 506--517.

\bibitem{neurostream}
E.~Azarkhish, D.~Rossi, I.~Loi, and L.~Benini, ``Neurostream: Scalable and
  energy efficient deep learning with smart memory cubes,'' \emph{IEEE
  Transactions on Parallel and Distributed Systems}, vol.~29, no.~2, pp.
  420--434, 2017.

\bibitem{space}
G.~Bertasius, H.~Wang, and L.~Torresani, ``Is space-time attention all you need
  for video understanding?'' in \emph{ICML}, 2021, p.~4.

\bibitem{google_workloads}
A.~Boroumand, S.~Ghose, Y.~Kim, R.~Ausavarungnirun, E.~Shiu, R.~Thakur, D.~Kim
  \emph{et~al.}, ``Google workloads for consumer devices: Mitigating data
  movement bottlenecks,'' in \emph{Proceedings of the Twenty-Third
  International Conference on Architectural Support for Programming Languages
  and Operating Systems}, 2018, pp. 316--331.

\bibitem{deep}
J.~Cai, M.~Takemoto, and H.~Nakajo, ``A deep look into logarithmic quantization
  of model parameters in neural networks,'' in \emph{Proceedings of the 10th
  International Conference on Advances in Information Technology (IAIT)}, 2018,
  pp. 1--8.

\bibitem{objectdetection}
N.~Carion, F.~Massa, G.~Synnaeve, N.~Usunier, A.~Kirillov, and S.~Zagoruyko,
  ``End-to-end object detection with transformers,'' in \emph{Computer
  Vision--ECCV 2020: 16th European Conference, Glasgow, UK, August 23--28,
  2020, Proceedings, Part I 16}.\hskip 1em plus 0.5em minus 0.4em\relax
  Springer, 2020, pp. 213--229.

\bibitem{eyeriss}
Y.-H. Chen, J.~Emer, and V.~Sze, ``Eyeriss: A spatial architecture for
  energy-efficient dataflow for convolutional neural networks,'' \emph{ACM
  SIGARCH computer architecture news}, vol.~44, no.~3, pp. 367--379, 2016.

\bibitem{HMC2_STD}
H.~M.~C. Consortium, ``Hybrid memory cube specification 2.1,'' 2014.

\bibitem{bert}
J.~Devlin, M.-W. Chang, K.~Lee, and K.~Toutanova, ``Bert: Pre-training of deep
  bidirectional transformers for language understanding,'' \emph{arXiv}, 2018.

\bibitem{image}
A.~Dosovitskiy, L.~Beyer, A.~Kolesnikov, D.~Weissenborn, X.~Zhai,
  T.~Unterthiner, M.~Dehghani, M.~Minderer, G.~Heigold, S.~Gelly \emph{et~al.},
  ``An image is worth 16x16 words: Transformers for image recognition at
  scale,'' \emph{arXiv}, 2020.

\bibitem{shidiannao}
Z.~Du, R.~Fasthuber, T.~Chen, P.~Ienne, L.~Li, T.~Luo, X.~Feng, Y.~Chen, and
  O.~Temam, ``Shidiannao: Shifting vision processing closer to the sensor,'' in
  \emph{Proceedings of the 42nd Annual International Symposium on Computer
  Architecture}, 2015, pp. 92--104.

\bibitem{tetris}
M.~Gao, J.~Pu, X.~Yang, M.~Horowitz, and C.~Kozyrakis, ``Tetris: Scalable and
  efficient neural network acceleration with 3d memory,'' in \emph{Proceedings
  of the Twenty-Second International Conference on Architectural Support for
  Programming Languages and Operating Systems}, 2017, pp. 751--764.

\bibitem{survey_accelerators}
D.~Ghimire, D.~Kil, and S.-h. Kim, ``A survey on efficient convolutional neural
  networks and hardware acceleration,'' \emph{Electronics}, vol.~11, no.~6, p.
  945, 2022.

\bibitem{quantization_survay}
A.~Gholami, S.~Kim, Z.~Dong, Z.~Yao, M.~W. Mahoney, and K.~Keutzer, ``A survey
  of quantization methods for efficient neural network inference,''
  \emph{arXiv}, 2021.

\bibitem{guo3d}
Q.~Guo, N.~Alachiotis, B.~Akin, F.~Sadi, G.~Xu, T.-M. Low, L.~Pileggi, J.~C.
  Hoe, and F.~Franchetti, ``3d-stacked memory-side acceleration: Accelerator
  and system design,'' 2014.

\bibitem{demystifying}
R.~Hadidi, B.~Asgari, B.~A. Mudassar, S.~Mukhopadhyay, S.~Yalamanchili, and
  H.~Kim, ``Demystifying the characteristics of 3d-stacked memories: A case
  study for hybrid memory cube,'' in \emph{IEEE international symposium on
  Workload characterization (IISWC)}, 2017, pp. 66--75.

\bibitem{oursurvey}
M.~Hassanpour, M.~Riera, and A.~Gonz{\'a}lez, ``A survey of near-data
  processing architectures for neural networks,'' \emph{Machine Learning and
  Knowledge Extraction}, vol.~4, no.~1, pp. 66--102, 2022.

\bibitem{long-tailed}
S.~Jain, S.~Venkataramani, V.~Srinivasan, J.~Choi, K.~Gopalakrishnan, and
  L.~Chang, ``Biscaled-dnn: Quantizing long-tailed datastructures with two
  scale factors for deep neural networks,'' in \emph{Proceedings of the 56th
  Annual Design Automation Conference 2019}, 2019, pp. 1--6.

\bibitem{new_DRAM_architecture}
J.~Jeddeloh and B.~Keeth, ``Hybrid memory cube new dram architecture increases
  density and performance,'' in \emph{IEEE Symposium on VLSI technology
  (VLSIT)}, 2012, pp. 87--88.

\bibitem{unison}
D.~Jevdjic, G.~H. Loh, C.~Kaynak, and B.~Falsafi, ``Unison cache: A scalable
  and effective die-stacked dram cache,'' in \emph{47th Annual IEEE/ACM
  International Symposium on Microarchitecture}, 2014, pp. 25--37.

\bibitem{HBM2}
H.~Jun, J.~Cho, K.~Lee, H.-Y. Son, K.~Kim, H.~Jin, and K.~Kim, ``Hbm (high
  bandwidth memory) dram technology and architecture,'' in \emph{2017 IEEE
  International Memory Workshop (IMW)}.\hskip 1em plus 0.5em minus 0.4em\relax
  IEEE, 2017, pp. 1--4.

\bibitem{log2_im2}
P.~Kareem, S.~R. Naqvi, and C.-M. Kyung, ``A low error add and shift-based
  efficient implementation of base-2 logarithm,'' in \emph{IEEE International
  Conference on Electrical Engineering (ICEE)}, 2017, pp. 1--6.

\bibitem{neurocube}
D.~Kim, J.~Kung, S.~Chai, S.~Yalamanchili, and S.~Mukhopadhyay, ``Neurocube: A
  programmable digital neuromorphic architecture with high-density 3d memory,''
  \emph{ACM SIGARCH Computer Architecture News}, vol.~44, no.~3, pp. 380--392,
  2016.

\bibitem{Samsung}
J.-S. Kim, C.~S. Oh, H.~Lee, D.~Lee, H.~R. Hwang \emph{et~al.}, ``A 1.2 v 12.8
  gb/s 2 gb mobile wide-i/o dram with 4 $\times$ 128 i/os using tsv based
  stacking,'' \emph{IEEE Journal of Solid-State Circuits}, vol.~47, no.~1, pp.
  107--116, 2012.

\bibitem{alexnet}
A.~Krizhevsky, I.~Sutskever, and G.~E. Hinton, ``Imagenet classification with
  deep convolutional neural networks,'' \emph{Communications of the ACM},
  vol.~60, no.~6, pp. 84--90, 2017.

\bibitem{understanding_dataflow}
H.~Kwon, P.~Chatarasi, M.~Pellauer, A.~Parashar, V.~Sarkar, and T.~Krishna,
  ``Understanding reuse, performance, and hardware cost of dnn dataflow: A
  data-centric approach,'' in \emph{Proceedings of the 52nd Annual IEEE/ACM
  International Symposium on Microarchitecture}, 2019, pp. 754--768.

\bibitem{maeri}
H.~Kwon, A.~Samajdar, and T.~Krishna, ``Maeri: Enabling flexible dataflow
  mapping over dnn accelerators via reconfigurable interconnects,'' \emph{ACM
  SIGPLAN Notices}, vol.~53, no.~2, pp. 461--475, 2018.

\bibitem{HBM}
D.~U. Lee, K.~W. Kim, K.~W. Kim, H.~Kim \emph{et~al.}, ``25.2 a 1.2v 8gb
  8-channel 128gb/s high-bandwidth memory (hbm) stacked dram with effective
  microbump i/o test methods using 29nm process and tsv,'' in \emph{IEEE
  International Solid-State Circuits Conference Digest of Technical Papers
  (ISSCC)}, 2014, pp. 432--433.

\bibitem{tsv}
J.~C. Lee, J.~Kim, K.~W. Kim, Y.~J. Ku, D.~S. Kim, C.~Jeong, T.~S. Yun, H.~Kim,
  H.~S. Cho, S.~Oh \emph{et~al.}, ``High bandwidth memory (hbm) with tsv
  technique,'' in \emph{IEEE International SoC Design Conference (ISOCC)},
  2016, pp. 181--182.

\bibitem{application}
V.~T. Lee, A.~Mazumdar, C.~C. del Mundo, A.~Alaghi, L.~Ceze, and M.~Oskin,
  ``Application codesign of near-data processing for similarity search,'' in
  \emph{IEEE International Parallel and Distributed Processing Symposium
  (IPDPS)}.\hskip 1em plus 0.5em minus 0.4em\relax IEEE, 2018, pp. 896--907.

\bibitem{dramsim3}
S.~Li, Z.~Yang, D.~Reddy, A.~Srivastava, and B.~Jacob, ``Dramsim3: A
  cycle-accurate, thermal-capable dram simulator,'' \emph{IEEE Computer
  Architecture Letters}, vol.~19, no.~2, pp. 106--109, 2020.

\bibitem{additive}
Y.~Li, X.~Dong, and W.~Wang, ``Additive powers-of-two quantization: An
  efficient non-uniform discretization for neural networks,'' in
  \emph{International Conference on Learning Representations (ICLR)}, 2020.

\bibitem{critical}
Z.~C. Lipton, J.~Berkowitz, and C.~Elkan, ``A critical review of recurrent
  neural networks for sequence learning,'' \emph{arXiv}, 2015.

\bibitem{processing}
J.~Liu, H.~Zhao, M.~A. Ogleari, D.~Li, and J.~Zhao, ``Processing-in-memory for
  energy-efficient neural network training: A heterogeneous approach,'' in
  \emph{51st Annual IEEE/ACM International Symposium on Microarchitecture
  (MICRO)}.\hskip 1em plus 0.5em minus 0.4em\relax IEEE, 2018, pp. 655--668.

\bibitem{log2_im1}
A.~Mansour, A.~El-Sawy, M.~Aziz, and A.~Sayed, ``A new hardware implementation
  of base 2 logarithm for fpga,'' \emph{International Journal of Signal
  Processing Systems}, vol.~3, no.~2, pp. 171--181, 2015.

\bibitem{recurrent}
T.~Mikolov, M.~Karafi{\'a}t, L.~Burget, J.~Cernock{\`y}, and S.~Khudanpur,
  ``Recurrent neural network based language model.'' in \emph{Interspeech},
  vol.~2, no.~3.\hskip 1em plus 0.5em minus 0.4em\relax Makuhari, 2010, pp.
  1045--1048.

\bibitem{convolutional}
D.~Miyashita, E.~H. Lee, and B.~Murmann, ``Convolutional neural networks using
  logarithmic data representation,'' \emph{arXiv}, 2016.

\bibitem{cacti}
N.~Muralimanohar, R.~Balasubramonian, and N.~P. Jouppi, ``Cacti 6.0: A tool to
  model large caches,'' \emph{HP laboratories}, vol.~27, p.~28, 2009.

\bibitem{scnn}
A.~Parashar, M.~Rhu, A.~Mukkara, A.~Puglielli, R.~Venkatesan, B.~Khailany,
  J.~Emer \emph{et~al.}, ``Scnn: An accelerator for compressed-sparse
  convolutional neural networks,'' \emph{ACM SIGARCH computer architecture
  news}, vol.~45, no.~2, pp. 27--40, 2017.

\bibitem{HMC}
J.~T. Pawlowski, ``Hybrid memory cube (hmc),'' in \emph{IEEE Hot Chips 23rd
  Symposium (HCS)}, 2011, pp. 1--24.

\bibitem{squad}
P.~Rajpurkar, J.~Zhang, K.~Lopyrev, and P.~Liang, ``Squad: 100,000+ questions
  for machine comprehension of text,'' \emph{arXiv}, 2016.

\bibitem{kernel_density}
S.~Seo and J.~Kim, ``Efficient weights quantization of convolutional neural
  networks using kernel density estimation based non-uniform quantizer,''
  \emph{Applied Sciences}, vol.~9, no.~12, p. 2559, 2019.

\bibitem{Designcompiler}
{Synopsys}, ``{Design Compiler},'' \url{https://www.synopsys.com/}, 2019.

\bibitem{ip}
\BIBentryALTinterwordspacing
Synopsys, ``Floating-point base-2 logarithm,'' 2023. [Online]. Available:
  \url{https://www.synopsys.com/dw/ipdir.php?c=DW_fp_log2}
\BIBentrySTDinterwordspacing

\bibitem{2023abndp}
B.~Tian, Q.~Chen, and M.~Gao, ``Abndp: Co-optimizing data access and load
  balance in near-data processing,'' in \emph{Proceedings of the 28th ACM
  International Conference on Architectural Support for Programming Languages
  and Operating Systems, Volume 3}, 2023, pp. 3--17.

\bibitem{L2L}
S.~Ullah, S.~Gupta, K.~Ahuja, A.~Tiwari, and A.~Kumar, ``L2l: A highly accurate
  log\_2\_lead quantization of pre-trained neural networks,'' in \emph{IEEE
  Design, Automation \& Test in Europe Conference \& Exhibition (DATE)}, 2020,
  pp. 979--982.

\bibitem{attention}
A.~Vaswani, N.~Shazeer, N.~Parmar, J.~Uszkoreit, L.~Jones, A.~N. Gomez,
  {\L}.~Kaiser, and I.~Polosukhin, ``Attention is all you need,''
  \emph{Advances in neural information processing systems}, vol.~30, 2017.

\bibitem{systematic_approach}
X.~Yang, J.~Pu, B.~B. Rister, N.~Bhagdikar, S.~Richardson, S.~Kvatinsky,
  J.~Ragan-Kelley, A.~Pedram, and M.~Horowitz, ``A systematic approach to
  blocking convolutional neural networks,'' \emph{arXiv}, 2016.

\bibitem{ptblm}
W.~Zaremba, I.~Sutskever, and O.~Vinyals, ``Recurrent neural network
  regularization,'' \emph{arXiv}, 2014.

\bibitem{transpim}
M.~Zhou, W.~Xu, J.~Kang, and T.~Rosing, ``Transpim: A memory-based acceleration
  via software-hardware co-design for transformer,'' in \emph{IEEE
  International Symposium on High-Performance Computer Architecture
  (HPCA)}.\hskip 1em plus 0.5em minus 0.4em\relax IEEE, 2022, pp. 1071--1085.

\end{thebibliography}
